\documentclass[12pt, a4paper]{article}
\usepackage[dvipdfmx]{graphicx}
\usepackage{amssymb}
\usepackage{amsmath}
\usepackage{bm}
\usepackage{color}
\usepackage{cite}
\usepackage{slashed}
\usepackage{subfigure}
\usepackage{epstopdf}            
\usepackage{epsfig}

\newcommand{\beq}{\begin{equation}}
\newcommand{\eeq}{\end{equation}}

\usepackage[colorlinks=true, linkcolor=black, citecolor=black,
urlcolor=blue]{hyperref} 

\begin{document}

\begin{titlepage}
\begin{center}

{\huge 
Probing $\mu\tau$ flavor-violating solutions for\\ the muon $g-2$ anomaly at Belle II}
\vskip 1.5cm
Syuhei Iguro$^1$, Yuji Omura$^{2}$, Michihisa Takeuchi$^{3}$
\vskip 0.5cm

{\it $^1$
Department of Physics, Nagoya University, Nagoya 464-8602, Japan}\\[3pt]

{\it $^2$ 
Department of Physics, Kindai University, Higashi-Osaka, Osaka 577-8502, Japan}\\[3pt]

{\it $^3$
Kobayashi-Maskawa Institute for the Origin of Particles and the
Universe, \\ Nagoya University, Nagoya 464-8602, Japan}\\[3pt]




\vskip 1.5cm

\begin{abstract}
The discrepancy between the measured value and the
Standard Model prediction of the muon anomalous magnetic moment is one of 
the most important issues in the particle physics. 
It is known that 
introducing a mediator boson X with the $\mu \tau$ lepton flavor violating (LFV) couplings is one good solution to explain the discrepancy, due to 
the $\tau$ mass enhancement in the one-loop correction.
In this paper, we study the signal of this model, i.e. the same-sign leptons,
in the Belle II experiment, assuming the flavor-diagonal couplings are suppressed. 
We show that the Belle II experiment is highly sensitive to 
the scenario in the mediator mass range of ${\cal O}(1-10)$~GeV, 
using the $e^+e^- \to \mu^\pm\tau^\mp X \to\mu^\pm \mu^\pm \tau^\mp \tau^\mp$ 
process induced by the $X$.\\
\end{abstract}
\end{center}
\end{titlepage}

\section{Introduction}
The longstanding discrepancy in the measured value of the anomalous magnetic moment of muon (muon $g-2$) and the SM prediction
might be a key to reveal the physics beyond standard model (BSM).
The estimated discrepancy between the experimentally measured value $a_\mu^{\rm{exp}}$ and
the SM prediction $a_\mu^{\rm{SM}}$ is currently given as~\cite{Blum:2018mom}\footnote{See also Refs.\cite{Hagiwara:2006jt,Jegerlehner:2009ry,Davier:2010nc,Hagiwara:2011af,Blum:2019ugy}. }
\begin{align}
\delta a_\mu=a_\mu^{\rm{exp}}-a_\mu^{\rm{SM}}=(2.74\pm0.73) \times10^{-9},
\label{g2muon}
\end{align}
and the significance reaches at 3.4~$\sigma$ level.
Currently a new experiment is in operation at the Fermilab (FNAL)~\cite{Grange:2015fou}  
and accumulating the data, which will reduce the uncertainty by a factor of four in the end of the planned operation.
Furthermore, another experiment is scheduled at the J-PARC in Japan~\cite{Mibe:2011zz}.
These experiments may confirm the discrepancy.
Motivated by this anomaly various models are proposed and tested so far\footnote{For a recent review, see, for example \cite{Lindner:2016bgg}.}.
Note that the discrepancy is at the same order of the electroweak contribution: $\delta a_\mu \simeq \delta a_\mu^{\rm EW} \simeq g^2m_\mu^2/16\pi^2m_W^2$.
The new physics contribution usually scales as $\delta a_\mu^{\rm NP} \simeq g_{\rm NP}^2 m_\mu^2/16\pi^2m_{\rm NP}^2$, 
where $g_{\rm NP}$ and $m_{\rm NP}$ are the new physics coupling and the mass. 
Up to now, there is no significant signal that suggests any particular new physics scenario found at the large hadron collider (LHC) 
nor at the various flavor experiments. 
The current situation may imply either the existence of the very heavy new particle to evade the LHC constraints 
with the corresponding large couplings, or the very light new particle with the corresponding small couplings. 
Already the current LHC bound is very severe, so that the former direction becomes less favored 
since the required coupling is too large to respect perturbativity.  

One well-known way to enhance new physics contribution keeping the coupling size small is
introducing a mediator with flavor violation couplings.
Since the dipole operator requires the chirality flipping,
one-loop correction involving a mediator with the $\mu\tau$ flavor violating couplings is enhanced
by a factor of $m_\tau/m_\mu \simeq {\cal O}(10)$.
Such $\mu\tau$ flavor violating scenarios are discussed to explain the muon $g-2$ anomaly in the context of 
the axion like particles (ALPs)~\cite{Bauer:2019gfk,Cornella:2019uxs}, the general two Higgs doublet model (G2HDM) \cite{Nie:1998dg,DiazRodolfo:2000yy,Iltan:2001nk,Wu:2001vq,Assamagan:2002kf,Davidson:2010xv,Omura:2015nja,Omura:2015xcg,Iguro:2018qzf,Abe:2019bkf,Iguro:2019sly,Wang:2019ngf}, 
and in the $Z^\prime$ models~\cite{Baek:2001kca,Heeck:2016xkh,Altmannshofer:2016brv,Iguro2020no1}.
\begin{figure}[b]
  \begin{center}
    \includegraphics[width=7cm]{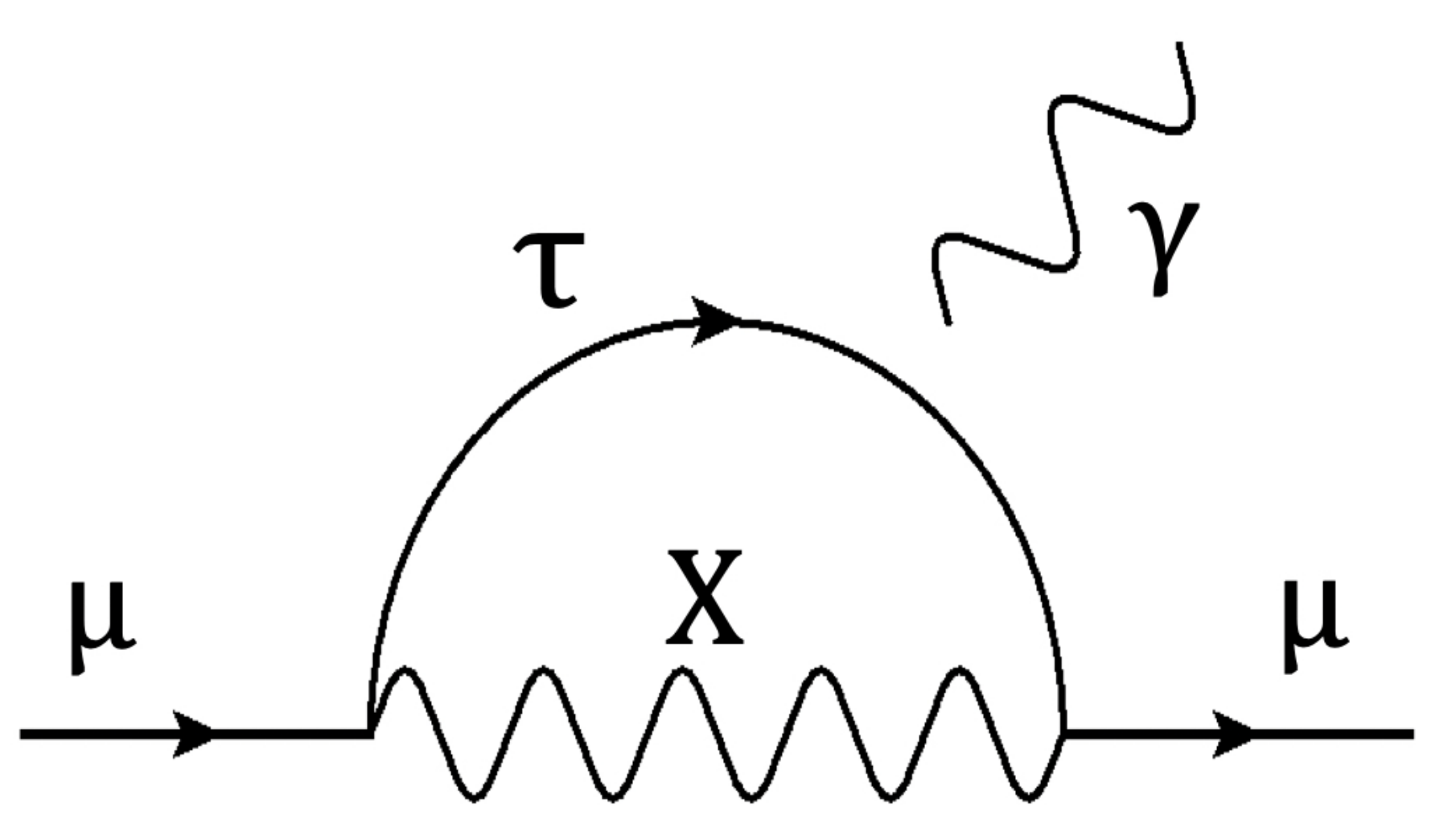}
            \caption{The figure shows the diagram to explain the anomaly for a $\mu\tau$ flavor violating X particle. We note that the internal fermion is a $\tau$ and this gives a so-called chirality enhancement because the chirality is flipped by the $\tau$ mass. }
    \label{diagramg-2}
  \end{center}
\end{figure}
Fig.~\ref{diagramg-2} shows the representative diagram to induce the $\delta a_\mu$ 
contribution with a $\mu\tau$ flavor violating mediator particle $X$, 
which is either a scalar or vector boson, and the $\mu\tau$ flavor violating nature of $X$ introduces the internal $\tau$-propagator at one-loop level.
It is important that $X$ has both left and right handed couplings to obtain the $m_\tau/m_\mu$ enhancement by chirality flipping at the internal $\tau$-propagator.

In general, lepton flavor violating (LFV) couplings are highly constrained by various low energy experiments. 
If there are flavor diagonal couplings of $X$ at the same time, it immediately induces large LFV effects; for instance, in $\tau \to \mu\gamma$, $\tau\to 3\mu $, $\tau\to \mu e e$, and so on.
On the other hand, if the interactions of $X$ are given by only $\mu\tau$ flavor violating couplings, testing the model in flavor experiments becomes difficult.
In this case, when the mass of the mediator $X$ is heavier than a few hundred GeV, the LHC is a powerful tool to test the scenario.
The authors of Ref. \cite{Altmannshofer:2016brv} have pointed out the importance of the searches for the events with 
the same sign muons and the opposite same sign taus ($\mu^\pm\mu^\pm\tau^\mp\tau^\mp$) 
to search for the $\mu\tau$ flavor violating vector bosons.
Recently, the importance of the photon initiated processes for the $\mu\tau$ flavor violating $Z^\prime$ searches is shown, 
and a large parameter region of the model favored by the anomaly would be covered at the high-luminosity (HL)-LHC~\cite{Iguro2020no1}.
Moreover, if the other neutral scalars exist as in G2HDMs,
the current LHC data for 150 fb$^{-1}$ would already cover the large parameter space 
through the electroweak production processes in the same $\mu^\pm\mu^\pm\tau^\mp\tau^\mp$ modes~\cite{Iguro:2019sly}.

On the other hand, when the mass is below $\mathcal{O}(10)$ GeV, it is rather difficult to search for such new particles at the LHC 
typically due to the limitation of the lepton detection 
for their transverse momentum below $10-20$ GeV~\cite{CMS:2017wua,Sirunyan:2018nwe}.
For instance, it is explicitly shown that the LHC is only sensitive down to $m_X=10$~GeV for such LFV particles~\cite{Altmannshofer:2016brv,Dev:2017ftk,Evans:2019xer}.
\begin{figure}[t]
  \begin{center}
     \includegraphics[width=8.3cm]{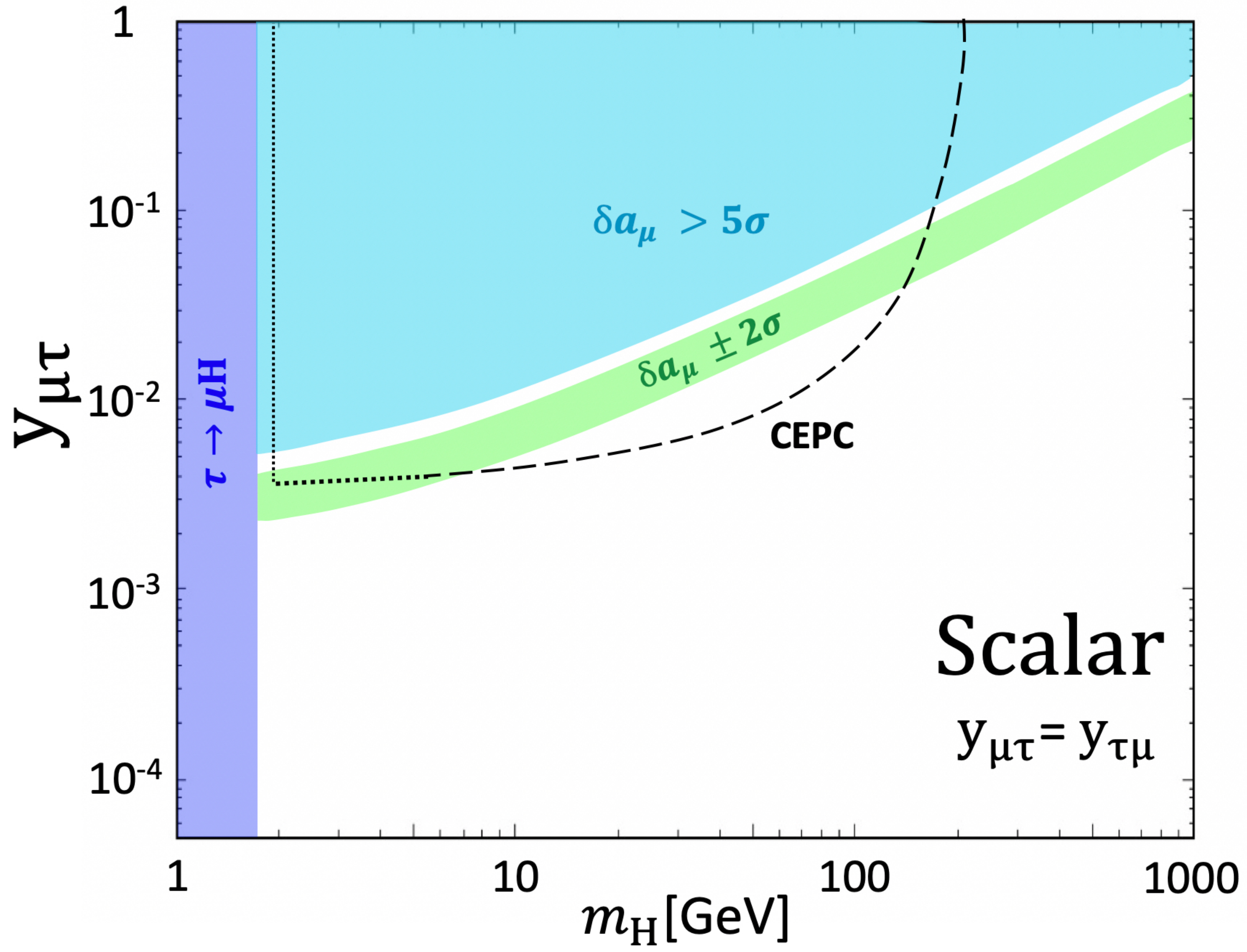}
    \includegraphics[width=8.45cm]{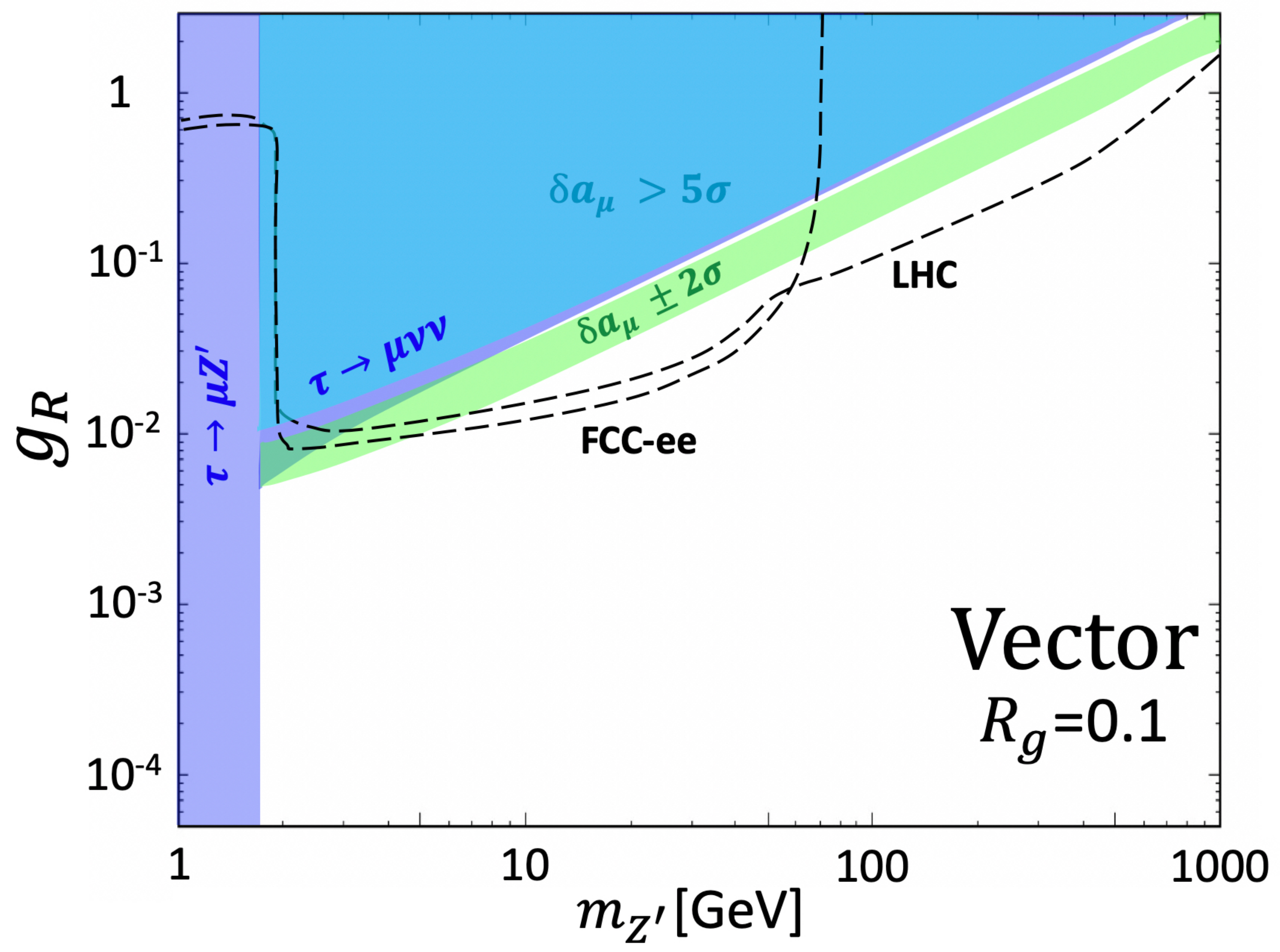}
            \caption{
            The summary of the current constraints and the future sensitivity for the scalar (left) and vector (right) $\mu\tau$ LFV  mediators 
            in the mass vs. coupling plane. The muon $g-2$ anomaly can be explained within $\pm2\sigma$ in green band, and the shaded regions in other colors are excluded.
The black lines show the prospects in the future collider experiments.
See the text for more details.
              }
    \label{fig:currentsituation}
  \end{center}
\end{figure}
Fig.~\ref{fig:currentsituation} summarizes the current allowed and the future testable regions 
in the mass vs. coupling plane. The left (right) panel shows the case 
for the scalar (vector) mediator case.
The horizontal axis represents the mass of the $\mu\tau$ LFV mediator ($m_{\rm{H}}$ and $m_{Z^{\prime}}$) for the case of scalar and vector, respectively). 
The vertical axis represents the corresponding  $\mu\tau$ LFV coupling, defined in Ref. \cite{Dev:2017ftk} for the scalar scenario, 
and defined in Sec. \ref{Vec}. For the scalar case, $y_{\mu\tau}=y_{\tau\mu}$ is assumed while $R_g=g_R/g_L=0.1$ for the vector case.

The green band shows the parameter space consistent with the discrepancy within $\pm2\sigma$. 
The blue shaded region shows the excluded region by the current constraint from $\tau$ decay data 
and the cyan shaded region are highly disfavored due to the too much contribution to $\delta a_\mu$ by more than 5$\sigma$.
The black dashed lines show the prospects for the future colliders, namely HL-LHC, FCC-ee 
for the vector scenario~\cite{Altmannshofer:2016brv,Iguro2020no1} 
and CEPC for the scalar scenario~\cite{Dev:2017ftk}.
Although Ref. \cite{Dev:2017ftk} shows the prospect only for $m_{\rm{H}} \ge$~5 GeV, 
we naively extrapolated the prospect down to $m_\tau+m_\mu$, indicated by the black dotted line.

From those summary plots, one can see that most of the parameter region is already constrained or testable at future colliders. 
However, the mass range of ${\cal O}(1-10)$~GeV would remain untestable even at future colliders,
 therefore, developing a new method to test the region is desired.
In this paper, we will show that the Belle II experiment~\cite{Abe:2010gxa,Kou:2018nap},
where the center of mass energy is $\sqrt {s_{\rm{BelleII}}}=10.58$ GeV, 
is especially sensitive to the mass range.
In this paper, we consider the mass range 
\begin{align}
m_\tau - m_\mu \le m_X \le \sqrt {s_{\rm{BelleII}}},
\label{massregion}
\end{align}
where $m_X$ is the mediator $X$ mass.
For $m_X \le m_\tau-m_\mu$, $X$ is copiously produced in $\tau$ decays, and 
a stringent constraint from $\tau\to \mu\nu\bar\nu$ ($\tau \to \mu~+$~invisible) would apply.
As a result, no parameter space is available to explain the muon $g-2$ anomaly~\cite{Tanabashi:2018oca}. 

\begin{figure}[t]
  \begin{center}
    \includegraphics[width=6cm]{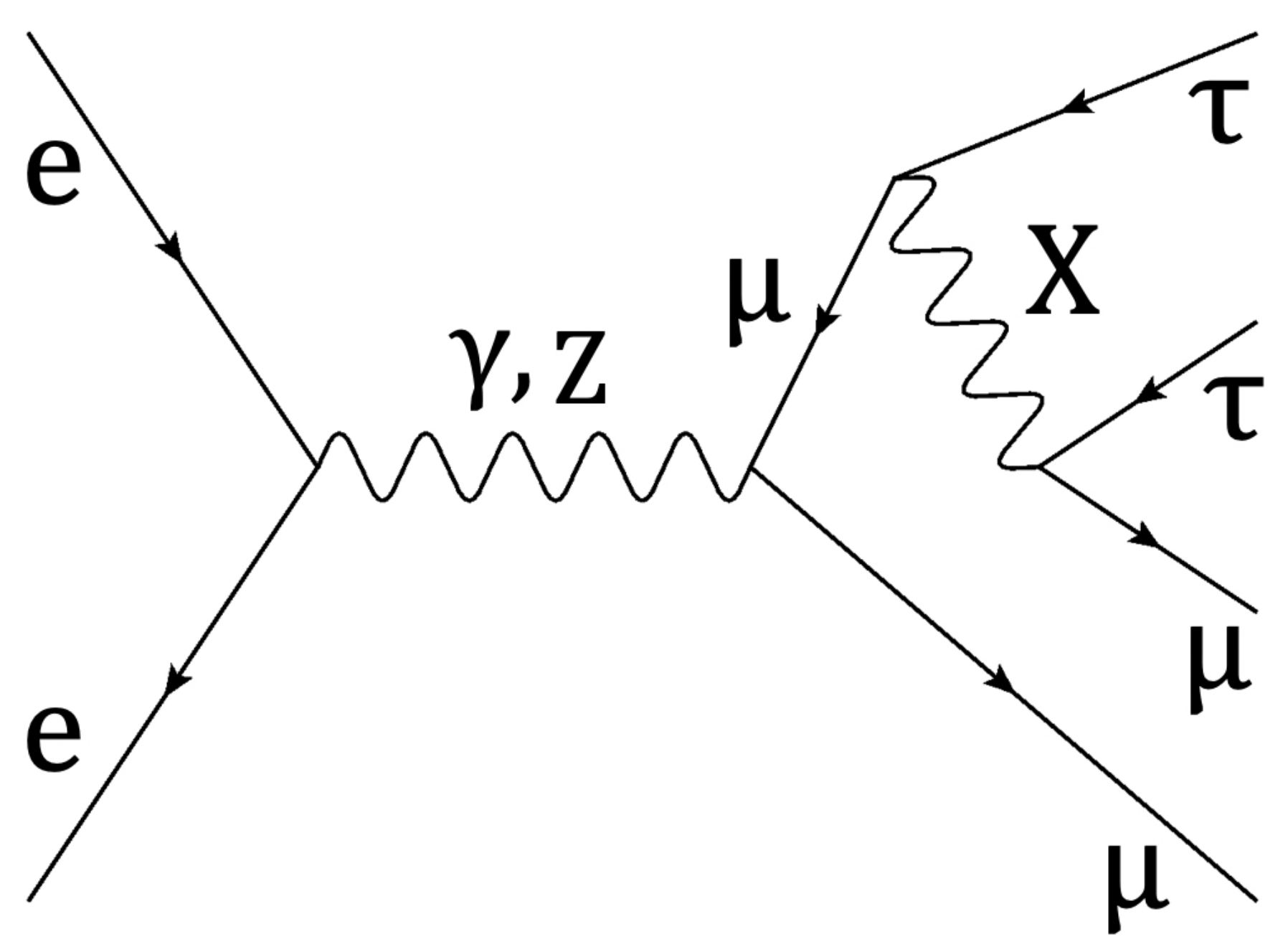}
            \caption{The relevant process at the Belle II experiments.
            There is also the one that obtained by exchanging $\mu$ and $\tau$.
            When a mediator X is light enough to be produced in the experiment, the cross section can be large.}
    \label{diagramBelle}
  \end{center}
\end{figure}

The mass region (\ref{massregion}) is further divided into two ranges.
In the mass range of  $m_\tau+m_\mu \le m_X \le  \sqrt {s_{\rm{BelleII}}}$, 
$X \to \mu\tau$ occurs as an on-shell process.
There we propose the search for the process: $e^+e^- \to \mu^\pm\tau^\mp X  \to \mu^\pm\mu^\pm\tau^\mp\tau^\mp$ as shown in Fig.~\ref{diagramBelle}\footnote{$e^+e^- \to e^\pm\mu^\mp a  \to e^\pm e^\pm \mu^\mp\mu^\mp$ is discussed where $e\mu$ flavor violating ALP (a) is introduced to explain the discrepancy in electron $g-2$\cite{Iguro2020no2}.}.
Unless $X$ is too heavy, they are produced at on-shell and cross section is expected to be not small. 

For $m_\tau-m_\mu \le m_X \le m_\tau+m_\mu$, $\tau \to \mu X$ is kinematically forbidden, 
therefore, no on-shell $X$ is produced in $\tau$ decays 
and evade the stringent bound from $\tau\to \mu\nu\bar\nu$ observation.
Since $m_X \le m_\tau+m_\mu$, 
unless $X$ couples to the lighter fermions or a pair of dark matter,
$X$ can not decay into 2-body, and can only undergo the 4-body decay through the off-shell $\tau$,
which makes $X$ a long lived particle.
Conservatively, $X$ can be treated as an invisible particle for both cases.
Nevertheless, the process $e^+e^- \to \mu^\pm\tau^\mp X$ (where $X$ is missing) would be sensitive, 
and a similar process discussed in Ref. \cite{Jho:2019cxq},
where the authors evaluated the sensitivity of $e^+e^- \to \mu^-\mu^+ Z^\prime$ (where $Z^\prime$ is missing) at the Belle II experiment.

This paper is organized as follows. In the Sec. 2 we introduce our simplified models to explain the muon $g-2$ anomaly, 
with a scalar mediator and with a vector mediator.
In Sec. 3, we study the signal at the Belle II experiment and show the potential coverage of the model parameters 
that can explain the anomaly. Sec. 4 is devoted to conclusion.

\section{{\bf{$\delta {a_\mu}$}} with a Scalar or Vector Particle}

In this section, we explain our phenomenological models where either scalar or vector mediator is introduced and the corresponding interaction is $\mu\tau$ flavor violating. 
In each model, we study the parameter region to explain the muon $g-2$ anomaly 
as well as relevant constraints.
We note that we focus on the mediator mass region below the invariant mass of the Belle II experiment: $m_X \le 10.58$~GeV.

\subsection{Scalar Scenario}
\label{Sca}
We discuss a model with a new scalar mediator.
Following the ALP scenario, we introduce the interaction as follows~\cite{Cornella:2019uxs}
\begin{align}
{\cal L_S}
&=-\frac{\partial_\mu a}{\Lambda}\sum_{i,j}^{\mu,\tau}\bar{l_i}\gamma^\mu (V^l_{ij}-a^l_{ij}\gamma_5)l_j \cr 
&=-i\frac{a}{\Lambda}\sum_{i,j}^{\mu,\tau}\bar{l_i}[(m_i-m_j)V^l_{ij}+(m_i+m_j)a^l_{ij}\gamma_5]l_j \ , 
\end{align}
where $a$ is ALP, that is a scalar mediator, and $V^l_{ij}$ ($a^l_{ij}$) are the vector (axial vector) couplings. 
$l_1$ and $l_2$ denotes the mass eigenstates of $\mu$ and $\tau$, so the off-diagonal elements of $V^l_{ij}$ and $a^l_{ij}$ correspond to the LFV couplings.
The scale $\Lambda$ is the cutoff scale of 
this effective Lagrangian, where a global symmetry is broken. Then, $a$ appears as a massless boson with the derivative coupling as in ${\cal L_S}$. 
We simply assume that $V^l_{ij}=V^l_{ji}$ and $a^l_{ij}=a^l_{ji}$ are satisfied and they are real.
In addition, the diagonal elements of $V^l_{ij}$ and $a^l_{ij}$ are vanishing in our setup.
Then, $V^l_{\mu\tau}$ and $a^l_{\mu\tau}$ are only relevant and the LFV processes are suppressed.
The scalar gains non-vanishing mass, $m_a$, as the ALP usually does, in our model.

In this setup, we have three independent parameters: the mass of the ALP particle $m_a$, 
scalar coupling $V^l_{\mu\tau}$, and pseudo scalar coupling $a^l_{\mu\tau}$. 
The $a_\mu$ contribution in this model is calculated as \cite{Cornella:2019uxs},
\begin{align}
\delta a_{\mu}^{{\cal S}}&=a_{\mu}^{{\cal S}}-a_{\mu}^{\rm{SM}} 
\simeq \frac{m_\mu^2}{16\pi^2\Lambda^2}\left[ \frac{m_\tau}{m_\mu}|V^l_{\mu\tau}|^2(1-|R_a|^2)\left(\frac{2x_{a}^2 \log x_{a}}{(x_{a}-1)^3}+\frac{1-3x_{a}}{(x_{a}-1)^2} \right) \right],
\end{align}
where $x_a=m_a^2/m_\tau^2$ and $R_a=a^l_{\mu\tau}/V^l_{\mu\tau}$. 
We note that the contribution vanishes in the limit of $|R_a|=1$, or $V^l_{\mu\tau}=\pm a^l_{\mu\tau}$.
Since the loop function is always positive, $|R_a|\le1$ needs to be satisfied to obtain 
the positive $\delta a_\mu^{\cal S}$ contribution. 
The smaller $|R_a|$, the smaller $|V^l_{\mu\tau}|$ is required to explain the anomaly.
Thus, we concentrate on the region with $0\le R_a\le1$ in the following\footnote{Even if we consider the case that $R_a$ is negative, our discussion does not change. }.

\begin{figure}[t]
  \begin{center}
    \includegraphics[width=6.6cm]{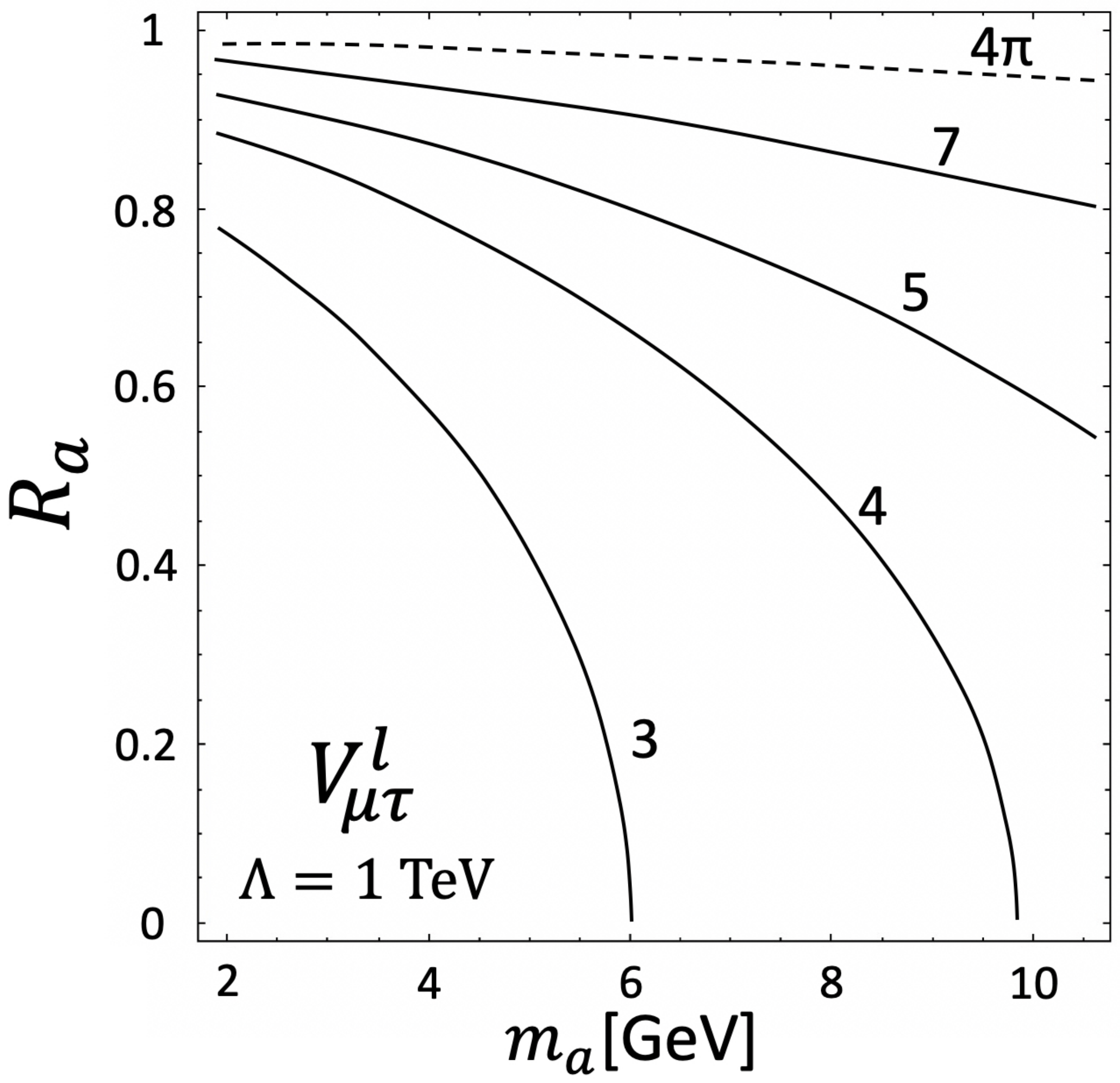}
  \includegraphics[width=7.5cm]{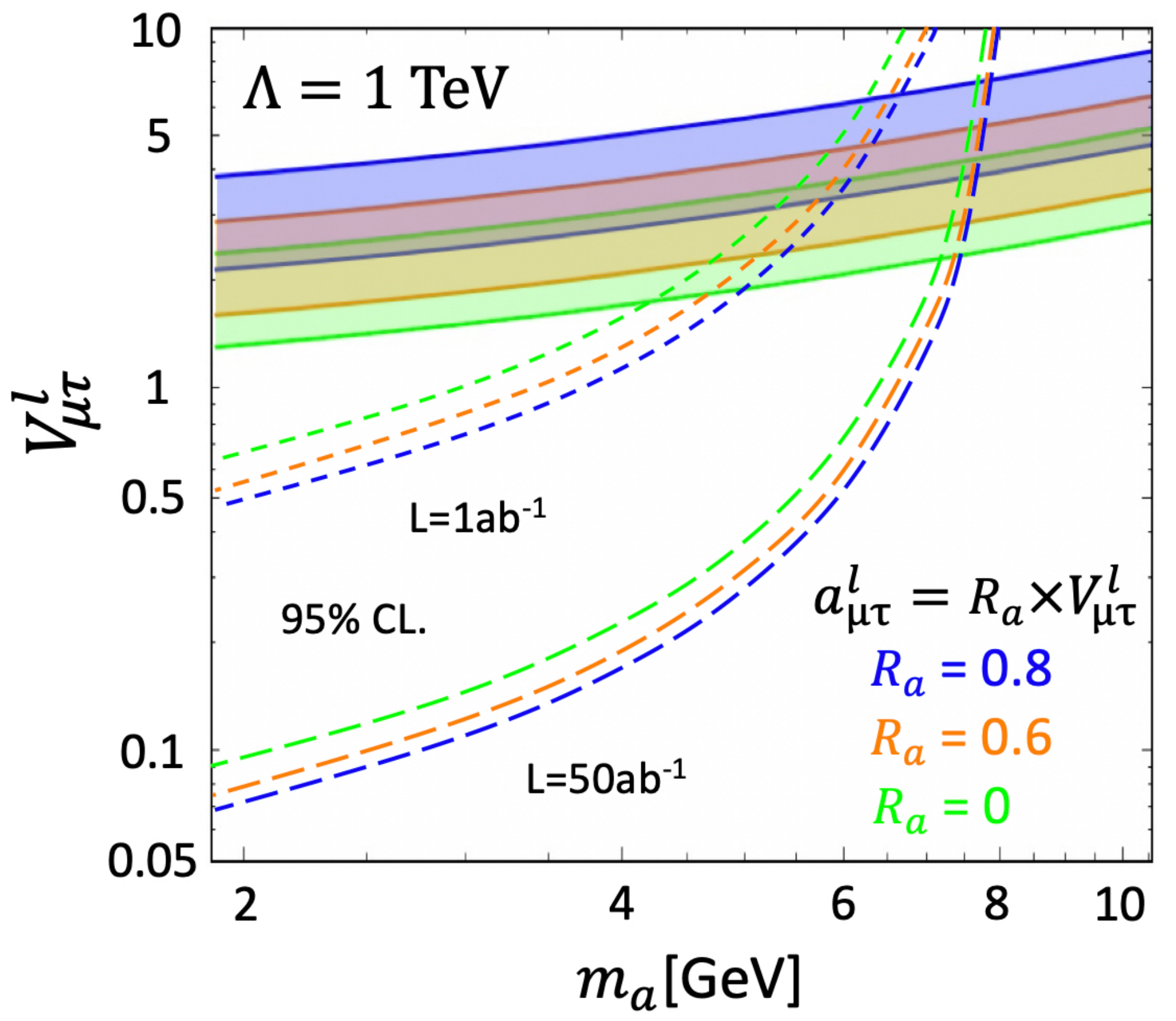}
             \caption{In the left figure, we show the value of $V^l_{\mu\tau}$, which is fixed to explain the central value of the anomaly in a mass vs. $R_a$ plane.
             In the right figure, we show the parameter space to explain the anomaly in a mass vs. $V^l_{\mu\tau}$ plane, where $\Lambda$ is fixed to be 1 TeV. 
             In each color bands the muon $g-2$ anomaly is explained within $\pm2\sigma$ where $R_a$ is fixed to be 0.8 (blue), 0.6 (orange) and 0 (green).
            The further description is given in the main text. 
            }
              \label{paramALP}
  \end{center}
\end{figure}
On the left panel in Fig.~\ref{paramALP}, $V^l_{\mu\tau}$ to obtain the central value for $\delta a_\mu$ is plotted on the $m_a$ vs. $R_a$ plane with $\Lambda=1$~TeV. 
The heavier $m_a$ region is disfavored by the perturbativity although the range we are considering in this paper would be acceptable for $\Lambda=1$~TeV.
On the right panel, the required value of $V^l_{\mu\tau}$ to explain the muon $g-2$ anomaly
as a function of $m_a$ is shown for each selected coupling ratio: $R_a=0.8$ (blue), 0.6 (orange), and 0 (green).  On the each band, the muon $g-2$ anomaly can be explained within $\pm2\sigma$.
As shown on the left panel, the larger $R_a$, the larger $V^l_{\mu\tau}$ is required.
On the right panel, $R_a$ is fixed at the same value on the each line as on the each band with the same color.  
The dashed (long-dashed) lines show our final estimate of the future prospect sensitivities by the Belle II experiment for the selected $R_a$ values with the integrated luminosity of 1 (50) ab$^{-1}$ at 95$\%$ confidence level (CL), that will be discussed in Sec.~\ref{Result}.
The larger $R_a$ is, the stronger the sensitivity becomes, 
since the production cross section is proportional to $(1+ R_a^2) (V^l_{\mu\tau})^2$,
but the contribution to $\delta a_\mu $ is proportional to $(1- R_a^2) (V^l_{\mu\tau})^2$.
Therefore, the limit $R_a=0$ would be the most difficult case to search for at the Belle II experiment.

\subsection{Vector Scenario}
\label{Vec}
Next, we consider another model with a vector mediator, $Z^\prime$.
If flavor-dependent gauged $U(1)$ symmetry is assigned to SM leptons, LFV gauge couplings of $Z^\prime$
are generally predicted. The texture of the couplings depend on the charge assignment and
the mass matrices for leptons. Now, we simple assume that 
the following LFV $Z^\prime$ couplings are effectively generated \cite{Altmannshofer:2016brv},
\begin{align}
\label{Zprime}
  {\cal L_{Z^\prime}}= g_L( \bar{\mu}_{L}\gamma^{\mu} \tau_{L}+ \bar{\nu}_{\mu L} \gamma^{\mu} {\nu_{\tau L}})Z^\prime_\mu+  g_R \bar{\mu}_{R}\gamma^{\mu} \tau_{R}Z^\prime_\mu+{\rm h.c.},
\end{align}
imposing SU(2)$_{\rm{L}}$ invariance.
We assume that the couplings $g_R$ and $g_L$ are real
and other $Z^\prime $ couplings are negligible. 
$Z^\prime $ gains non-vanishing mass, $m_{Z^\prime}$, according to the spontaneous symmetry breaking.
Thus, this model consists of the three free parameters: $Z^\prime$, $m_{Z^\prime}$, $g_L$, and $g_R$.

\begin{figure}[t]
  \begin{center}
             \includegraphics[width=5.5cm]{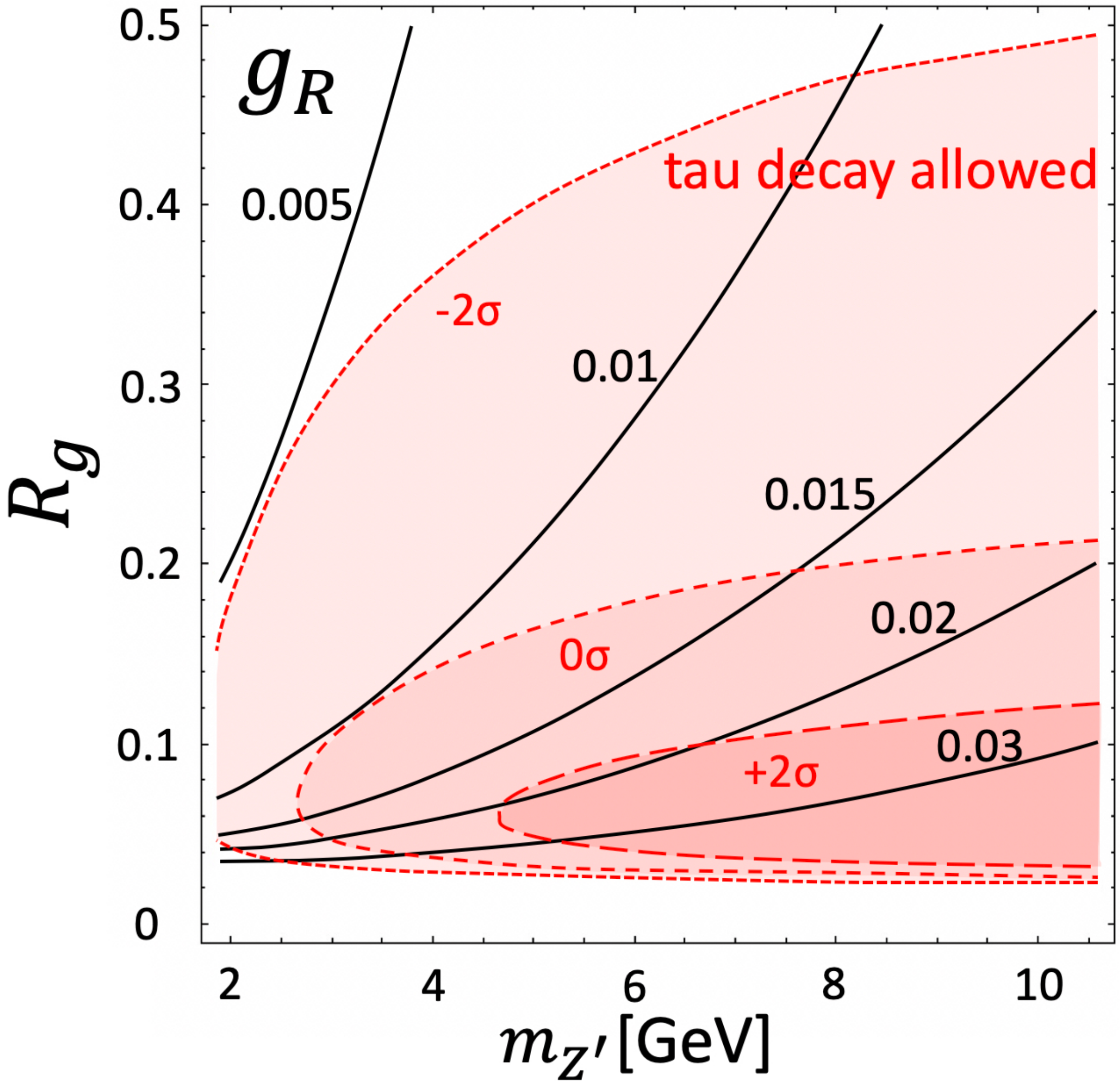}
            \includegraphics[width=6.05cm]{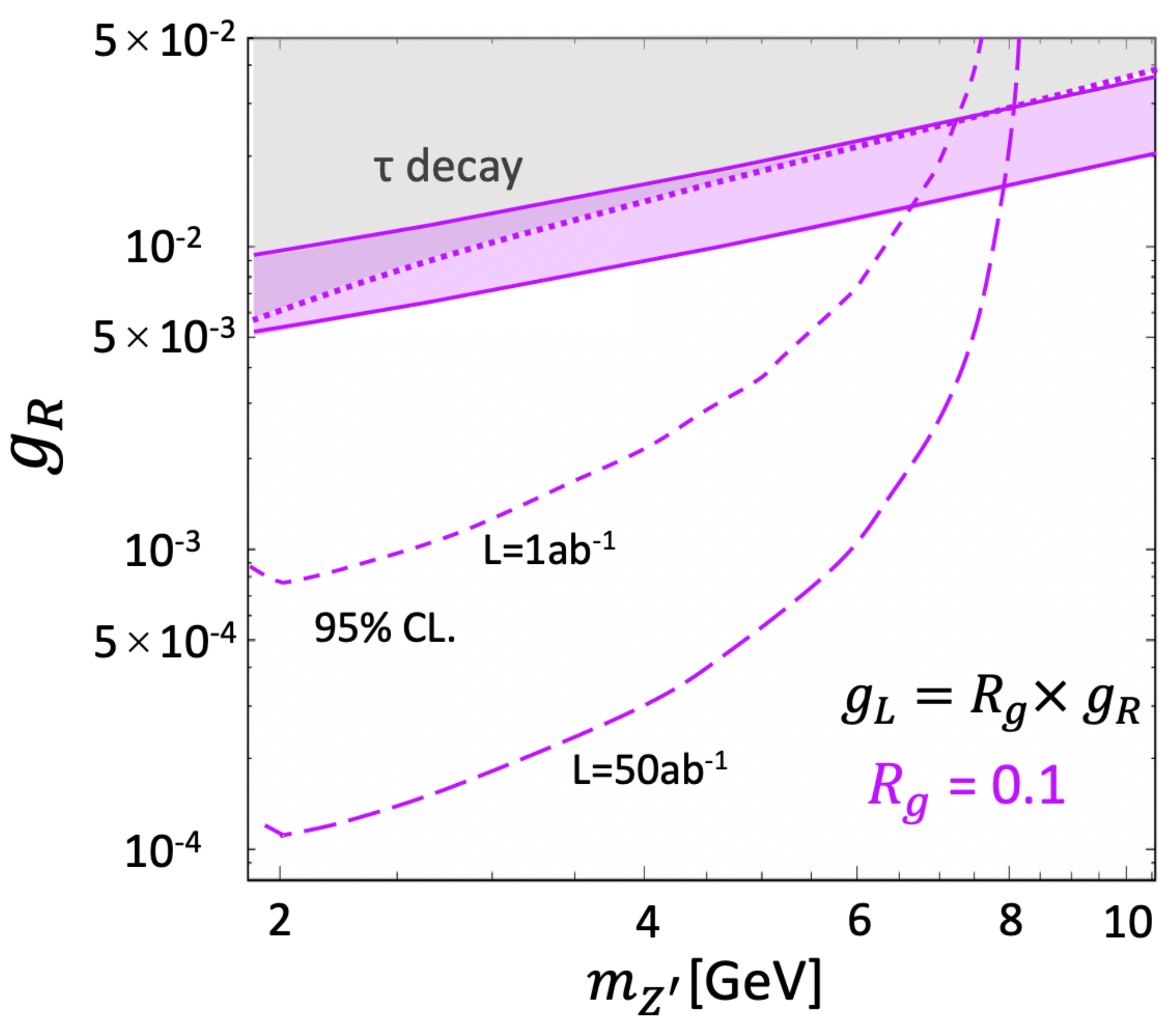}
        \includegraphics[width=5.95cm]{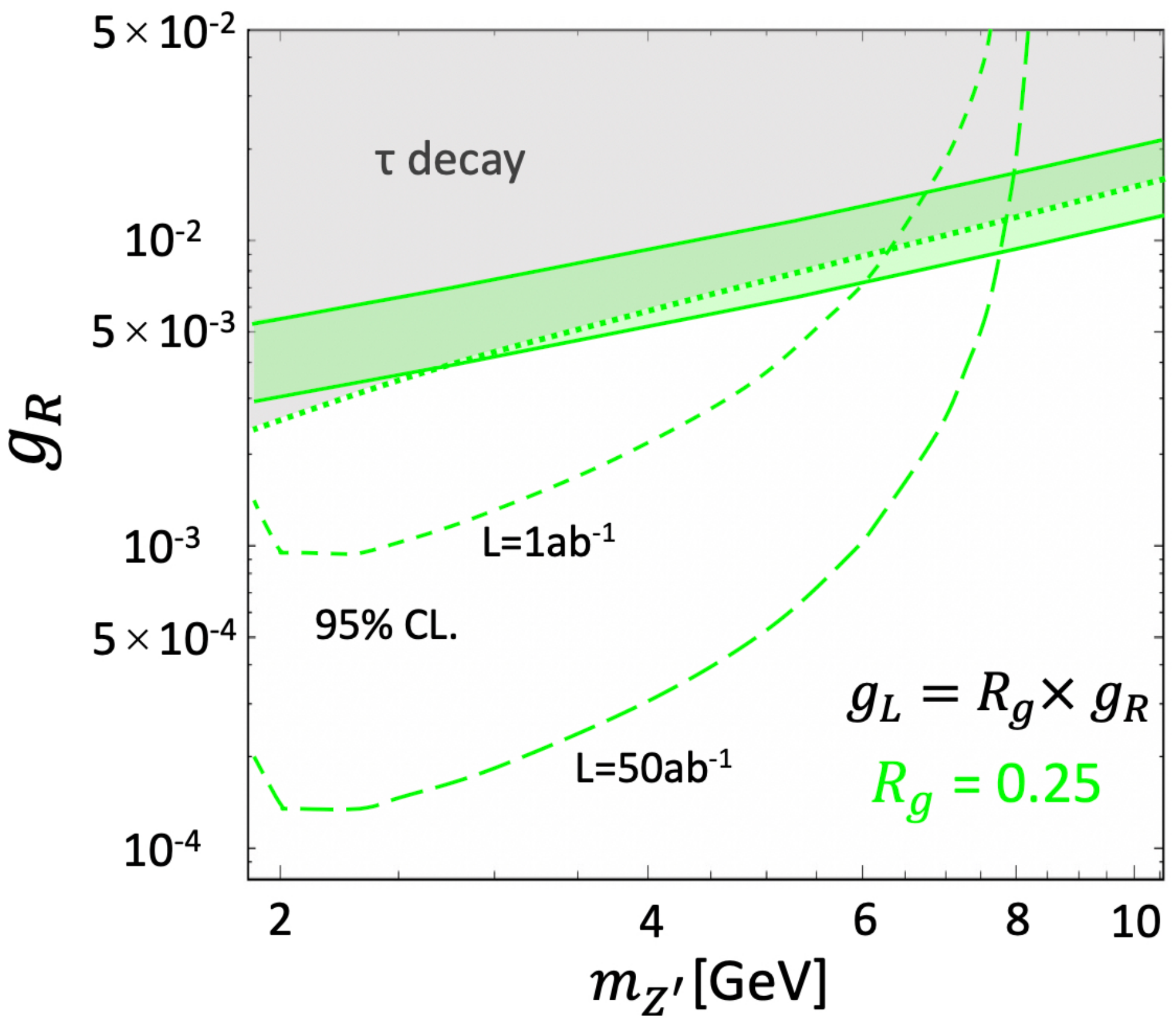}
    \includegraphics[width=6.05cm]{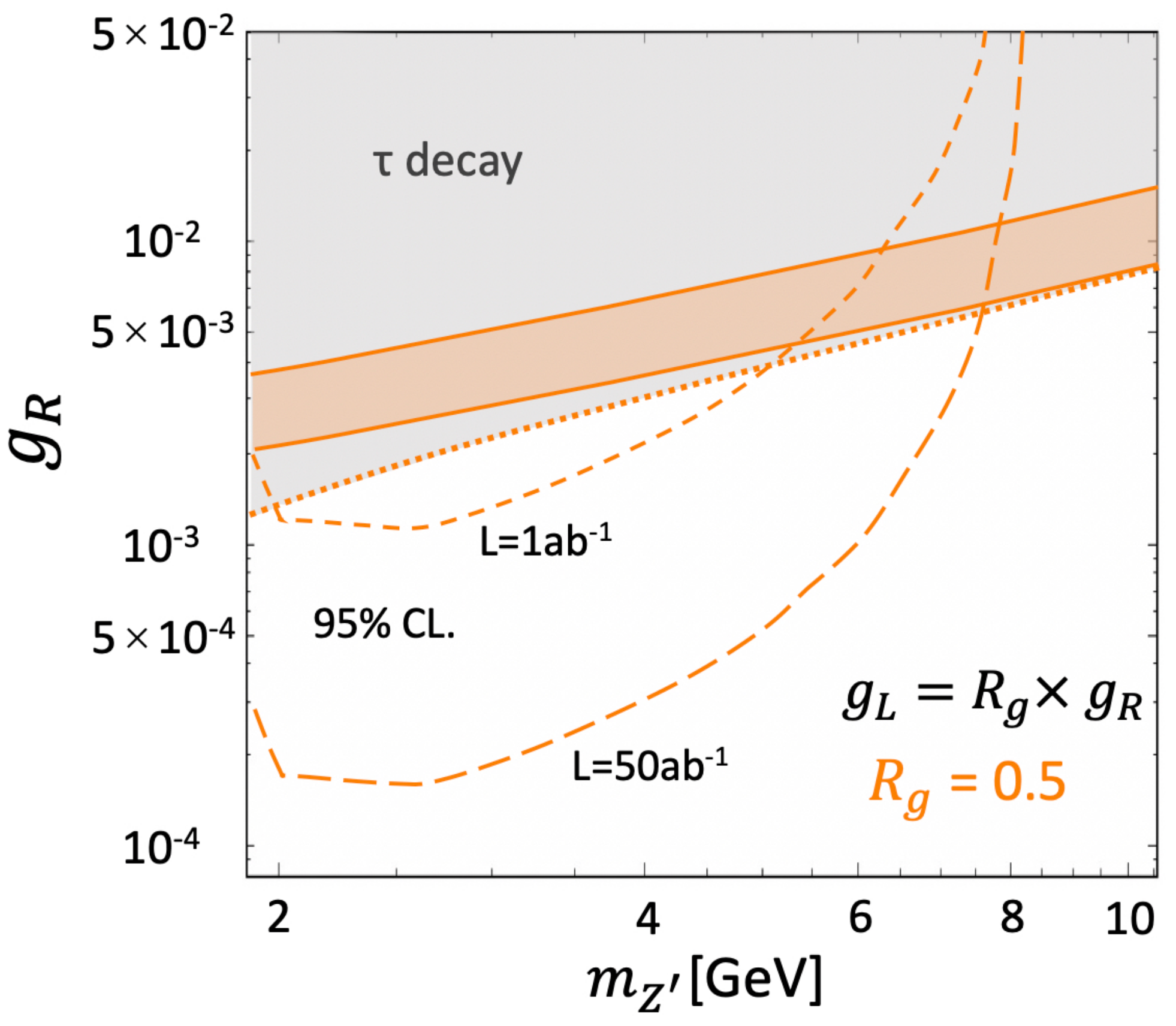}
            \caption{The top left panel of Fig.~\ref{paramZp} shows the contour plot for $g_R$ required to explain the anomaly in the $R_g$ vs. $m_{Z^\prime}$ plane. 
             The model  can explain the anomaly with -2$\sigma$, 0$\sigma$ and +2$\sigma$, the lightest red shaded region, middle red shaded region, red shaded region are allowed, respectively.   
            We also show the parameter region that explains the anomaly for $R_g=$ 0.1 (purple, left lower), 0.25 (orange, right above), and 0.5 (orange, left above) in the $g_R$ vs. $m_{Z^\prime}$ plane.
            See the main text for the more detailed description.
 }  
    \label{paramZp}
  \end{center}
\end{figure}

In this model, the contribution to the muon $g-2$ is given as, 
\begin{align}
\delta a_{\mu}^{\cal {Z^\prime}}&=\frac{m_\mu^2}{16\pi^2}  \int_0^1 dx g_R^2\biggl[ 2(1+R_g^2)\left( (1-x) \left(\left( x^2-2x\right)+\frac{x^2m_\tau^2}{2m_{Z^\prime}^2} \right) \right)\nonumber\\
&~~~~~~~~~~~~~~~~~~+4 R_g  \frac{m_\tau}{m_\mu} \left(2(x-x^2) +\frac{x^2m_\tau^2}{2m_{Z^\prime}^2}\right)\biggl]\times\biggl[m_{Z^\prime}^2(1-x)+xm_\tau^2\biggl]^{-1},
\end{align} 
where we define $R_g=g_L/g_R$.
The loop function proportional to $(1+R_g^2)$ provides the negative contribution, 
while the one proportional to $R_g$ provides the positive contribution in the mass region we are interested in.
Thus, $R_g$ has to be positive to explain the $a_\mu$ anomaly. 
The top left panel of Fig.~\ref{paramZp} shows the contour plot of the $g_R$ required to 
explain the anomaly on the $R_g$ vs. $m_{Z^\prime}$ plane.
When the model explains the anomaly with -2$\sigma$, 0$\sigma$ and +2$\sigma$, the lightest red shaded region, middle red shaded region, red shaded region are allowed.   
The numbers along black lines correspond to the value of $g_R$ to achieve the central value of $\delta a_\mu$. 
When one requires that the model explains the anomaly within -2$\sigma$ (+2$\sigma$) level,
the required $g_R$ values get smaller (larger) by a factor of 0.68 (1.24).
The tau decay $\tau\to\mu\nu\nu$ is also affected by the existence of such a light $Z^\prime$. 
On the top left panel in Fig.~\ref{paramZp}, the regions allowed by the tau decay constraints are indicated in red color. The three regions correspond to the cases where one requires the $\delta a_\mu$ value to be the central value, and the $\pm 2 \sigma$ values respectively. As one can see, 
we could obtain the lower bound on $m_{Z^\prime}$ and the upper bound on $R_g$. 
For example, $m_{Z^\prime} \ge 2.7$ GeV is required to explain the central value.
To obtain the $\delta a_\mu$ within the $2\sigma$ value, $0.02<R_g<0.5$ is required. 
A full mass range relevant to the Belle II is still available when $R_g=0.1$.
Depending on the magnitude of the coupling, we have the different parameter region allowed by the $\tau $ decay constraint.
This is because the leading contribution to the $\tau\to\mu\nu\nu$ comes from the 
interference term between the SM and the $Z^\prime$ amplitude (Re[$M_{\rm{SM}}M_{Z^\prime}^*]$) and is proportional to $g_L^2$.
When $R_a$ is going down to $\mathcal{O}(0.01)$, the $|M_{Z^\prime}|^2$ term, which is proportional to $g_R^2 g_L^2$, becomes the relevant contribution.

We also show the $2\sigma$ regions on $m_{Z^\prime}$ vs. $g_R$ plane 
in Fig.~\ref{paramZp}.
On the top right panel, $R_g$ is fixed at $R_g=$ 0.1.
On the bottom left and right, $R_g$ is fixed at $R_g=$ 0.25 and 0.5, respectively.
The grey shaded region indicates the excluded region by the tau decay constraints at 95 $\%$ CL. for 
each $R_g$ value~\cite{Tanabashi:2018oca}.
For each $R_g$ value, the dashed (long-dashed) line shows our estimate of the future prospect 95$\%$ CL. sensitivity 
at the Belle II experiment with 1 (50) ab$^{-1}$ of the integrated luminosity, that will be discussed in Sec.~\ref{Result}.
We will see that the smaller the $R_g$ is, the larger cross section is predicted. 
Therefore, $R_g=0.5$ case is most difficult case to exclude by the Belle II experiment. Still in that case about up to $m_{Z^\prime}$=6 (8) GeV would be excluded at 1 (50) ab$^{-1}$ of the data.

\section{Collider Signals at the Belle II experiment}
\label{Coll}
In this section, we estimate the sensitivity of the models at the Belle II experiment.
The Belle II experiment is in operation at the SuperKEKB collider, which is an asymmetric $e^+ e^-$ collider with 
$E_{e^+}=4$~GeV and $E_{e^-}=7$ GeV, corresponding to the center of the energy of 10.58~GeV.
The final planned integrated luminosity is $\int {\rm L} d{\rm t}$ = 50 ab$^{-1}$~\cite{Kou:2018nap}.
We focus on the searches only using the following 4-lepton process, 
\begin{align}
e^+e^-\to \mu^\pm\tau^\mp X\to \mu^\pm \mu^\pm \tau^\mp \tau^\mp \ {\rm or}\ \mu^\pm \mu^\mp \tau^\pm \tau^\mp,
\end{align}
As $X$ can decay into $\mu^\pm \tau^\mp$ independent of the sign of the intermediate step, we should have 
the same number of $\mu^\pm\mu^\pm \tau^\mp \tau^\mp$ and $\mu^\pm\mu^\mp \tau^\pm \tau^\mp$ events.
While the former process is essentially the SM background (BG) free, the SM processes can contribute to the latter process.
For simplicity, we only consider the former process $\mu^\pm\mu^\pm \tau^\mp \tau^\mp$ 
in this paper, and inclusion of the latter processes is beyond the scope of 
the paper although it would improve the sensitivity further.
Thus our sensitivity estimate 
should be taken as a conservative one.
The distinctive features of our signal events are as follows:
\begin{itemize}
\item{Two pairs of the same sign muons and the same sign taus, with each pair has the opposite sign}
\item{One pair of $\mu$ and $\tau$ forms the resonance of $X$.}
\end{itemize}
The both features are distinctive and useful to distinguish the signal from the BG 
and we simply assume the BG is negligible after the selection of the signs 
at the first approximation\footnote{We note that the charge identification of tau leptons 
is also good at the Belle II experiment since the charges of the decay products can be reconstructed 
with a very good accuracy~\cite{Hanagaki:2001fz,Ikado:2006un,Kou:2018nap}.}.
We note that ALP and $Z^{\prime}$ decays immediately for the nearly entire mass region even if we take a boost factor of the Belle experiment into account.
When ALP mass is very slightly larger than the sum of muon and tau masses, the lifetime can be long because of the phase space suppression.
In this case, the experimental signature will be $\mu^\pm\tau^\mp$+missing and can be a target at the Belle II
experiment.
\subsection{Signal and Kinematic Cut}
We generate the signal events 
using {\tt MadGraph5}~\cite{Alwall:2014hca} and {\tt PYTHIA8}~\cite{Sjostrand:2006za} interfaced
with the model file generated by {\tt FeynRules}~\cite{Alloul:2013bka}, 
and estimate the signal cross section.
To guarantee the tau identification we do not include the muonic decaying taus as a signal.
Different from at the high energy colliders, e.g. LHC, taus are not such boosted at the Belle II due to its 
low center of mass energy.
Thus, not all the flying directions of the $\tau$ decay products are collimated in the original $\tau$ direction, 
and the fact that the $\tau$ direction is in the acceptance region of the detector does 
not guarantee that all the visible decay products are found in the detector.
To obtain a conservative estimate of the acceptance and the efficiency for the decaying taus, we require
all visible decay products from the taus and the prompt muons satisfy the following set of the experimental kinematic cuts
at the laboratory frame~\cite{Hanagaki:2001fz,Kou:2018nap}, and multiplied the conservative efficiency $\epsilon=0.9$ for each object, where $\theta_i$ is the angle between the $e^-$ direction and the direction of the particle $i$, and $\vec{p}_{i}$ is its three momentum, respectively.
\begin{itemize}
\item{for $\mu$ : $25^\circ \le \theta_{\mu} \le 145^\circ, ~|\vec{p}_\mu|\ge0.6\rm{~GeV}$,}
\item{for $e$ : $12^\circ \le \theta_{e} \le 155^\circ, ~|\vec{p}_e|\ge0.02\rm{~GeV}$,}
\item{for $\pi^\pm$, $\pi^0$ and $K^\pm$: $12^\circ \le \theta_{\pi, K} \le 150^\circ,~ |\vec{p}_{\pi,K}|\ge0.02\rm{~GeV}$,}
\end{itemize}  

For example, if the $\tau$ decays into $\pi^+\pi^- \pi^+ \pi^0$+ missing, we require the all acceptance cut and applies 
the $\epsilon^4 = 0.9^4$ for the efficiency factor.
The multiplicity distribution for the whole signal events 
peaks around 6, which would already end up with $\epsilon^6 \simeq 0.53$. 
The resulting factor including the acceptance and the efficiency 
for the aforementioned $\mu^\pm\mu^\pm \tau^\mp \tau^\mp$ events is in general 
about 25\% for most of the cases. 

\subsection{Sensitivity}
\label{Result}
\begin{figure}[t]
  \begin{center}
    \includegraphics[width=8.35cm]{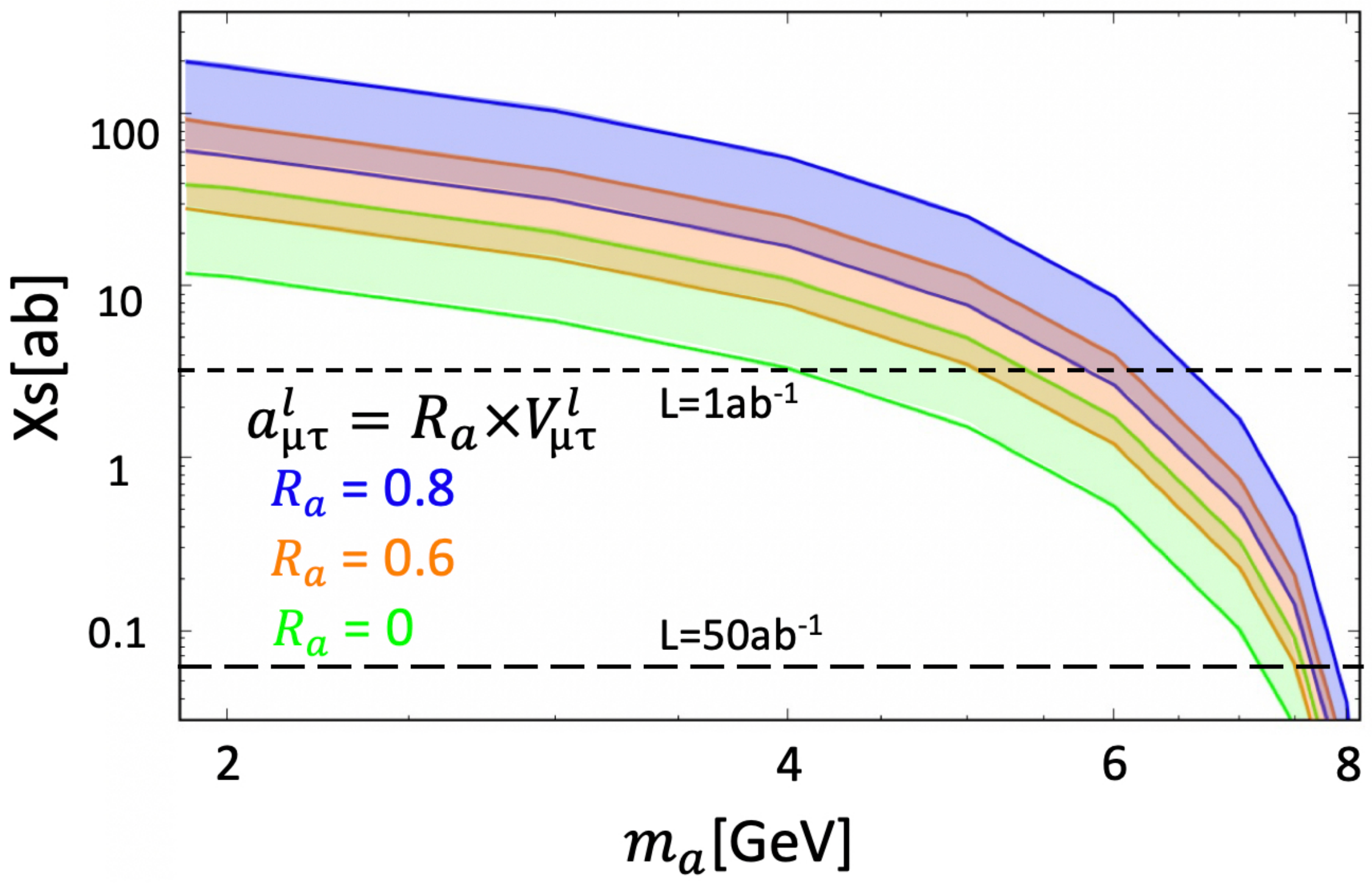}
    \includegraphics[width=8.35cm]{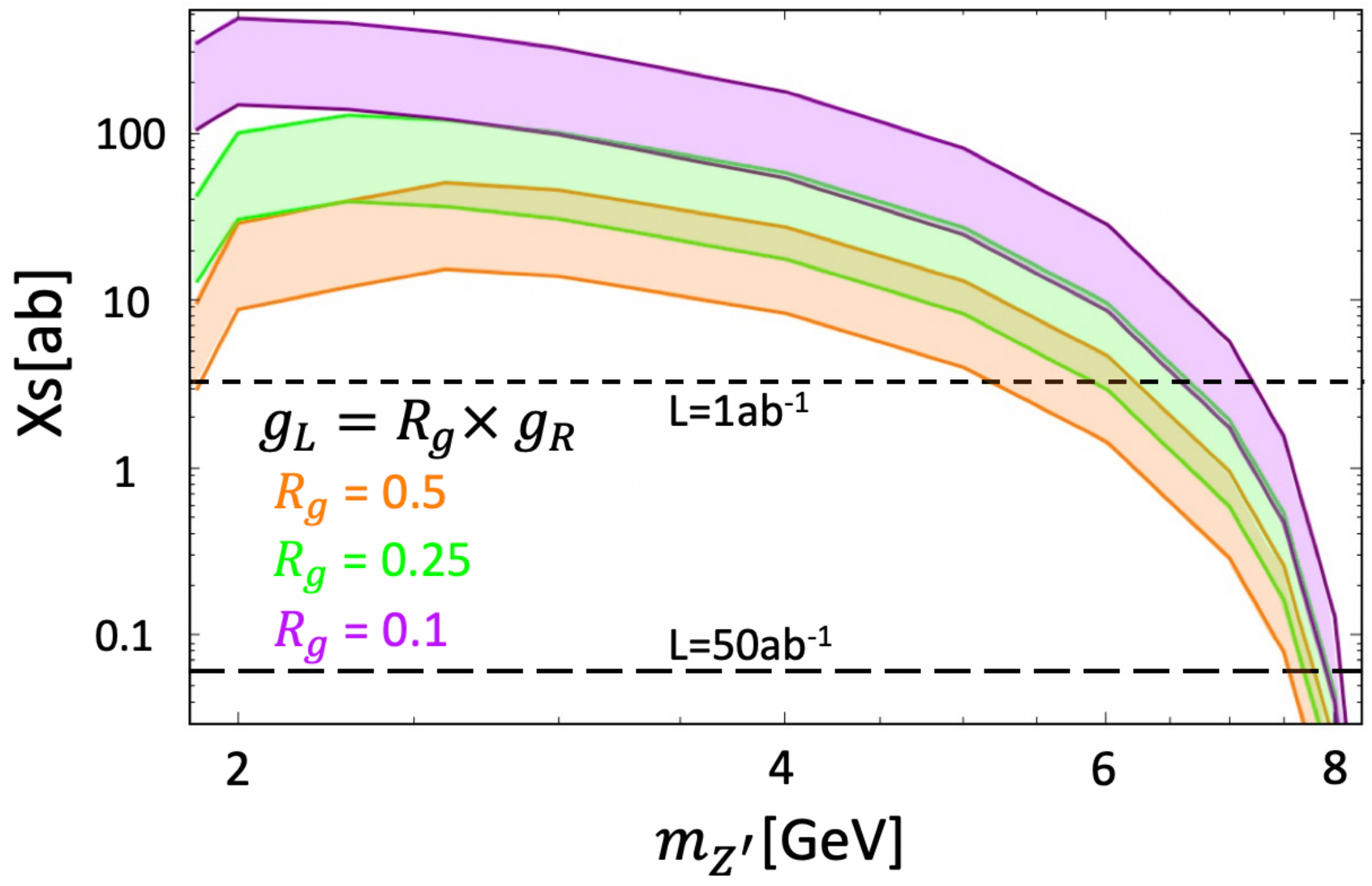}
            \caption{The left figure shows the fiducial cross section (the vertical axis) of $e^+e^-\to \mu^\pm\mu^\pm\tau^\mp\tau^\mp$ process after the cut in ab against the mass (the horizontal axis) for ALP.
Light green, orange and blue bands are prediction for $R_a$=0, $R_a$=0.6 and $R_a$=0.8 within $\pm2\sigma$ of the anomaly, respectively. 
The fiducial cross section of $e^+e^-\to \mu^\pm\mu^\pm\tau^\mp\tau^\mp$ after the kinematic cut in ab vs. mass of $Z^\prime$ is shown in the right figure.
             Light green, orange and purple bands are prediction for $R_g$=0.5, $R_g$=0.25 and $R_g$=0.1, respectively. 
             We plot the future prospect at Belle II experiment in black dashed (long dashed) line for 1 (50) ab$^{-1}$ of data.
              The region above the line will be tested.}
    \label{result}
  \end{center}
\end{figure}

We show the prediction of the fiducial cross section (Xs) in a unit of ab
after imposing the acceptance cuts and the efficiencies for the scaler (ALP) scenarios (left panel) 
and for the vector scenarios (right panel) in Fig.~\ref{result}. 
The number of events for 50 ab$^{-1}$ of the data is simply obtained by multiplying a factor of $50$ to the Xs.

On the left panel, blue, orange, and light green bands describe the fiducial cross sections for $R_a=0.8$, $0.6$ and $0$, respectively. The smaller the $R_a$ is, the smaller the $V_{\mu\tau}$ is required to explain the anomaly, which corresponds to the smaller fiducial cross section.
For each color bands, the upper and the lower boundaries are corresponding to the cases accommodating 
the $a_\mu$ anomaly at the significance of $\pm2 \sigma$.
Since we have a small number of signal events with no background, we need to use Poisson statistics.
The exclusion at 95$\%$ confidence level is given when 3.09 signal events are predicted, when no events are observed experimentally \cite{Tanabashi:2018oca}.
The horizontal dotted black lines show the experimental sensitivity at 1 ab$^{-1}$ (upper) and 50 ab$^{-1}$ (lower), and the region above the lines can be tested.
As the $m_a$ approaches to the kinematical limit of $\sqrt{s_{\rm Belle}} - (m_\tau + m_\mu) \simeq 8.7~$GeV, 
the fiducial cross section rapidly drops due to the phase space suppression.
Still the large part of the surviving parameter region favored by the muon $g-2$ anomaly will be covered at the Belle II.
For instance, we find that 1 (50) ab$^{-1}$ of the data can test up to 6 (7.5) GeV in the $R_a=0.8$ scenario.
However, for $R_a=0$ the minimum coupling size is required and the corresponding cross section is minimized, 1 (50) ab$^{-1}$ of the data can test up to 4 (7) GeV.
We also interpret the results in terms of the coupling vs. mass plane as in Fig.~\ref{paramALP}.

\begin{figure}[t]
  \begin{center}
     \includegraphics[width=8.3cm]{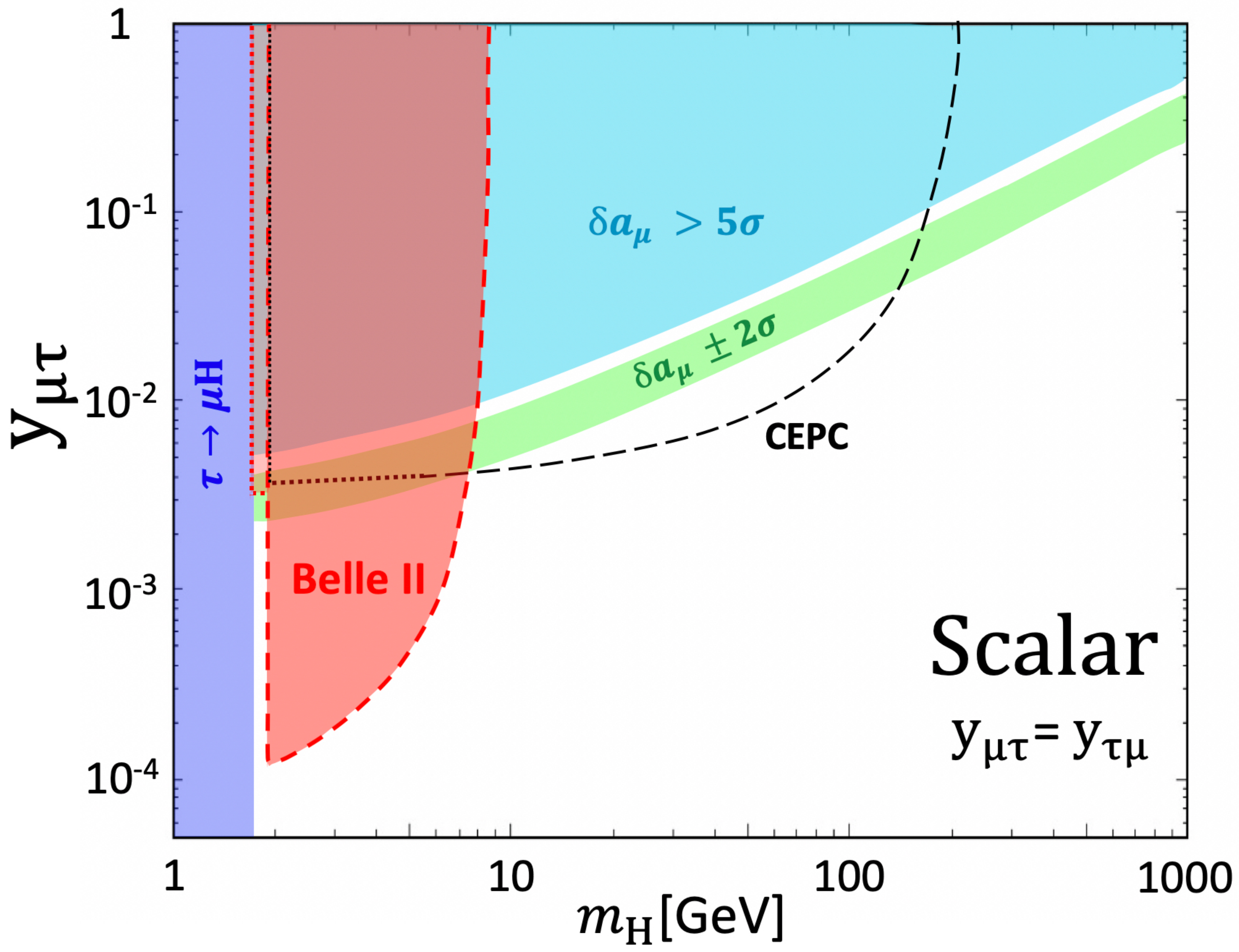}
    \includegraphics[width=8.45cm]{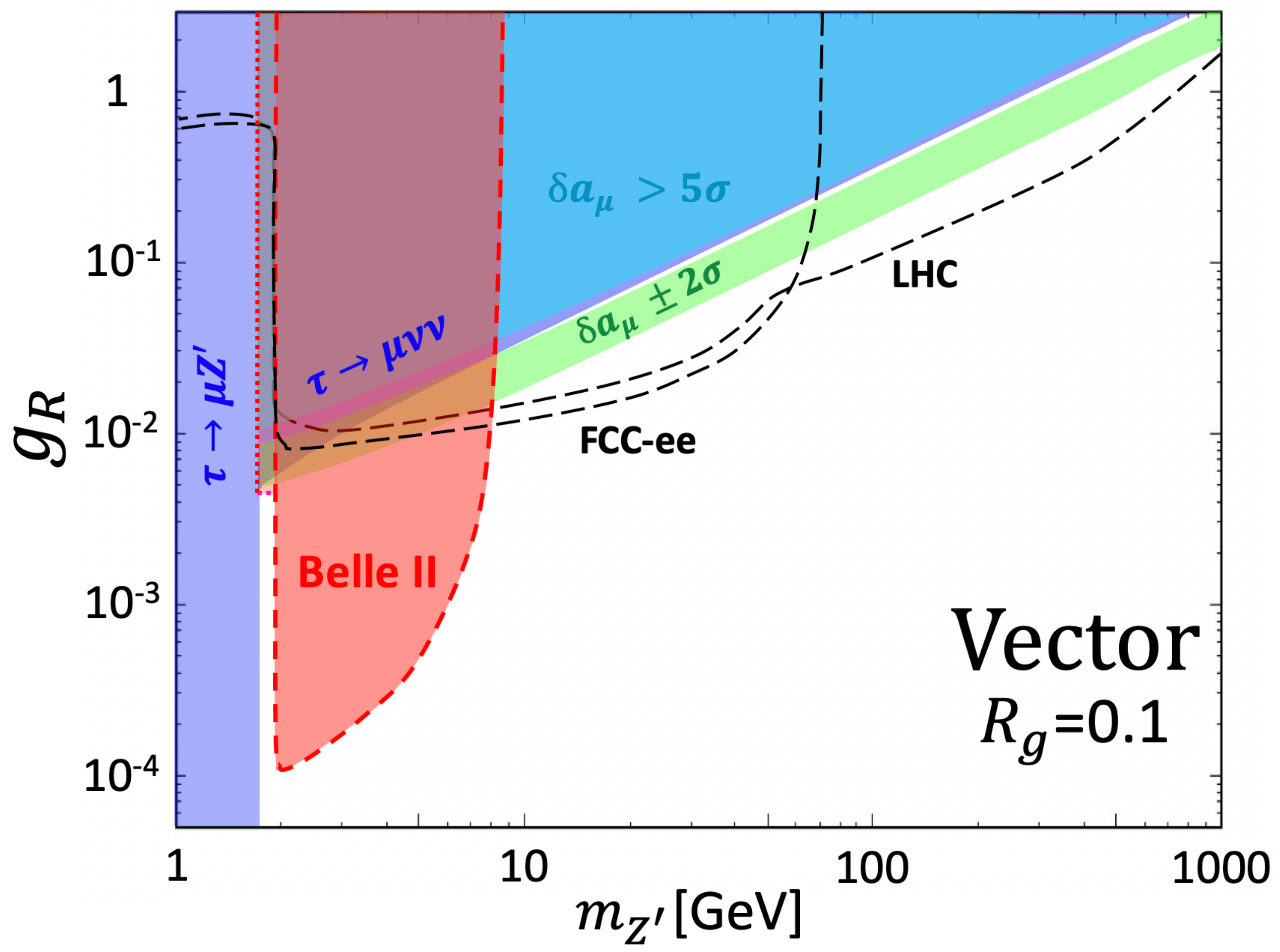}
            \caption{The future prospect for a $\mu\tau$ FLV scalar boson (left) and $\mu\tau$ FLV vector boson (right) including the Belle II projection are summarized. The red long dashed line and dotted line are prospect for the $e^+e^-\to \mu^\pm \mu^\pm\tau^\mp \tau^\mp $ process and  $e\mu$+missing search, respectively. The red shaded region can be tested with 50 ab$^{-1}$ of the data. The other information is the same as for Fig.~\ref{fig:currentsituation}.
              }
    \label{fig:future}
  \end{center}
\end{figure}

On the right panel in Fig.~\ref{result}, we show the corresponding results for the $Z^\prime$ models for $R_g=0.1$ (purple), 0.25 (light green), and 0.5 (orange), respectively.
Note that the number of $Z^\prime \to \mu\tau$ events is suppressed due to the phase space suppression 
when $m_{Z^\prime}$ approaches to $m_\tau+m_\mu$, under the existence of 
the unsuppressed $Z^\prime \to \nu\bar{\nu}$ contributions. 
The BR($Z^\prime\to\mu\tau$) is expressed as
$(1+\Gamma_{Z^\prime\to\nu\nu}/\Gamma_{Z^\prime\to\mu\tau})^{-1}$, where the partial width ratio is given as
\begin{align}
\frac{\Gamma_{Z^\prime\to\nu\nu}}{\Gamma_{Z^\prime\to\mu\tau}}&=\frac{2R_g^2}{\sqrt{\left(1-\frac{(m_\tau+m_\mu)^2}{m_{Z^\prime}^2}\right) \left(1-\frac{(m_\tau-m_\mu)^2}{m_{Z^\prime}^2}\right)} \left((1+R_g^2)\left(2+\frac{m_\tau^2+m_\mu^2}{m_{Z^\prime}^2}-\frac{\left(m_\tau^2-m_\mu^2\right)^2}{m_{Z^\prime}^4}\right)+3R_g\right)}.
\end{align}
As a result the fiducial cross section decreases in the light mass region especially for the large $R_g$ scenario.
For large mass region, the cross section drops toward the kinematic limit $m_{Z^\prime} \simeq 8.7$~GeV 
by the same reason as the scalar case. The horizontal dotted lines show the 95\% CL. sensitivity 
for the corresponding integrated luminosity assuming no background contributes. 
For instance, we found that 1 (50) ab$^{-1}$ of the data can test up to 6.5 (7.5) GeV 
in the $R_g=0.1$ scenario. Even with the sensitivity drop at the smaller $m_{Z^\prime}$ region, 
the integrated luminosity of 1 ab$^{-1}$ would cover the range down to $m_{Z^\prime} \simeq m_\tau+m_\mu$
assuming the anomaly is explained within $\pm2 \sigma$.
Similarly to the scaler case, we can again obtain the upper bound of the coupling $g_R$ as a function of $m_{Z^\prime}$ from the efficiency and the acceptance, which we show in Fig.~\ref{paramZp}.
Thus the large mass region can be tested at Belle II.
Finally, Fig.~\ref{fig:future} shows our Belle II sensitivity with the red long dashed lines overlaid to the summary 
plots quoted as Fig.~\ref{fig:currentsituation}.

As discussed above, $X$ can not decay into $\mu\tau$ when $m_X \le m_\tau+m_\mu$. 
In that case, and the dominant decay mode becomes the 4 body decay $X \to \mu^+ \mu^- \nu_\mu \nu_\tau$, or $X \to \nu_\mu \nu_\tau$ if it exists. Since the 4 body decay easily becomes long-lived for the detector, and anyway $X$ is observed as a missing particle, 
they provide $e^+e^-\to \mu^\pm\tau^\mp$+missing events at Belle II~\footnote{If $X$ decays in the detector, even more exotic events would be observed and it would be easier to detect.}.
To estimate the sensitivity in the mass region, 
we follow the Belle II analysis searching for the $e^+e^-\to e^\pm \mu^\mp$+missing events~\cite{Adachi:2019otg}.
Note that the $\tau^\mp$ subsequently decays into $e^\pm\nu\bar \nu$, 
and  $\mu^\pm\tau^\mp$+missing events eventually provide the $\mu^\pm e^\mp$+missing events.
We have estimated the efficiency to find the recoil mass $\le$ 7 GeV after imposing the several acceptance cuts for our signals, 
and found 13 $\%$ (14 $\%$) for the scalar (vector) scenario on top of the standard $BR(\tau \to e \nu\bar\nu)=17\%$.
On the other hand,  the fiducial cross section for the SM BG is at most about 15~fb 
according to Fig.~4 in the reference.
From those numbers, taking the criteria of $S/\sqrt{B}=2$ as the definition of the future sensitivity, 
we found $Y_{\mu\tau}=3.3\times10^{-3}$  ($g_R=4.7\times10^{-3}$) would be sensitive 
for the scalar (vector) scenario at Belle II with 50 ab$^{-1}$.
They are shown by a red dotted line in Fig.~\ref{fig:future} in the range of $m_\tau-m_\mu\le m_X\le m_\tau+m_\mu$. 
Note that the signal to background ratio is extremely small for such a small couplings, 
which is $S/B \sim 0.04\rm{fb}/15\rm{fb}=3\times10^{-3}$
because the kinematic cuts in the reference is very minimal.
Since a large number of signals $S \sim 2000$ is expected, we expect there is an enough room for 
the more dedicated study to improve it. The other $\tau$ decay mode would also help to improve the sensitivity.

We note that if the mediators couples to dark matter particles $\chi$, the signal and the sensitivity 
would be reduced due to the additional $X \to \chi\chi$ modes. 
In that case, we have to consider also the contributions from $\mu^\pm\tau^\mp$+missing mode. 
To discuss this possibility the model needs to be specified and we leave this possibility for the future work. 


\section{Conclusion and Discussion}

The discrepancy between the experimental measurements and the theoretical predictions 
observed in the muon anomalous magnetic moment may imply the existence of new physics.
Although various models have been proposed and tested so far, 
no other explicit convincing signals have been observed in any experiments.
In this paper, we discussed the models with a scalar and vector mediator with $\mu\tau$ flavor violating couplings,
and explicitly show that relatively small coupling can explain the discrepancy in both models 
due to the chirality enhancement by a factor of $m_\tau/m_\mu$ in the muon $g-2$ contribution. 

We have shown that both models can evade the various current experimental constraints and a large parameter space is still available. Moreover, we have also shown that the various future experiments 
will probe the most of the parameter region which can explain the current 
discrepancy of the muon $g-2$, while only the region of the mass range of $\mathcal{O}(1-10)$ GeV will remain untestable 
using the proposals in the literatures. 
In this paper, we have proposed the way to search for the mediators in the range of $\mathcal{O}(1-10)$ GeV 
at the Belle II experiment, using the distinctive signature of $\mu^\pm\mu^\pm\tau^\mp\tau^\mp$.
We estimated the sensitivity at the Belle II experiment and found that the large part of the parameter is testable with the data of $\mathcal{O}(10)$ ab$^{-1}$. The number we provided is only using the most distinctive signature $\mu^\pm\mu^\pm\tau^\mp\tau^\mp$ and should be taken as a conservative estimate. The more dedicated study by the experimentalist would be desired.
Once the signature is found, it would not be difficult to discriminate the scalar and vector models 
using the angular distributions.

\section*{Acknowledgements}
The authors thank C.-P. Yuan, Zhite Yu, Kirtimaan~A.~Mohan, Kazuhiro Tobe, Motoi Endo, Teppei Kitahara, Kenji Inami, Kodai Matsuoka, Gianluca Inguglia, and Ilya Komarov for valuable discussions. 
The work of S. I. is supported by Kobayashi-Maskawa Institute for the Origin of Particles and the Universe, Toyoaki scholarship foundation and the Japan Society for the Promotion of Science (JSPS) Research Fellowships for Young Scientists, No. 19J10980.
It is pleasure for S. I. to thank Michigan State University for its warm hospitality where this work was started. 
The work of Y. O. is supported by Grant-in-Aid for Scientific research from the Ministry of Education, Science, Sports, and Culture (MEXT), Japan, No. 19H04614, No. 19H05101, and No. 19K03867.
M. T. is supported in part by the JSPS Grant-in-Aid for Scientific Research
No.~16H03991, 16H02176, 18K03611, and 19H04613.



\begin{thebibliography}{99}
{\small{
\bibitem{Blum:2018mom} 
  T.~Blum {\it et al.} [RBC and UKQCD Collaborations],
  Phys.\ Rev.\ Lett.\  {\bf 121}, no. 2, 022003 (2018)
  [arXiv:1801.07224 [hep-lat]].
  
\bibitem{Hagiwara:2006jt} 
  K.~Hagiwara, A.~D.~Martin, D.~Nomura and T.~Teubner,
  Phys.\ Lett.\ B {\bf 649}, 173 (2007)
  [hep-ph/0611102].

\bibitem{Jegerlehner:2009ry} 
  F.~Jegerlehner and A.~Nyffeler,
  Phys.\ Rept.\  {\bf 477}, 1 (2009)
  [arXiv:0902.3360 [hep-ph]].

\bibitem{Davier:2010nc} 
  M.~Davier, A.~Hoecker, B.~Malaescu and Z.~Zhang,
  Eur.\ Phys.\ J.\ C {\bf 71}, 1515 (2011)
  [arXiv:1010.4180 [hep-ph]].

\bibitem{Hagiwara:2011af}
  K.~Hagiwara, R.~Liao, A.~D.~Martin, D.~Nomura and T.~Teubner,
  J.\ Phys.\ G {\bf 38} (2011) 085003
  [arXiv:1105.3149 [hep-ph]].

\bibitem{Blum:2019ugy} 
  T.~Blum, N.~Christ, M.~Hayakawa, T.~Izubuchi, L.~Jin, C.~Jung and C.~Lehner,
  arXiv:1911.08123 [hep-lat].

\bibitem{Grange:2015fou} 
  J.~Grange {\it et al.} [Muon g-2 Collaboration],
  arXiv:1501.06858 [physics.ins-det].
  
\bibitem{Mibe:2011zz} 
  T.~Mibe [J-PARC g-2 Collaboration],
  Nucl.\ Phys.\ Proc.\ Suppl.\  {\bf 218}, 242 (2011).

\bibitem{Lindner:2016bgg} 
  M.~Lindner, M.~Platscher and F.~S.~Queiroz,
  Phys.\ Rept.\  {\bf 731}, 1 (2018)
  [arXiv:1610.06587 [hep-ph]].

\bibitem{Bauer:2019gfk} 
  M.~Bauer, M.~Neubert, S.~Renner, M.~Schnubel and A.~Thamm,
  arXiv:1908.00008 [hep-ph].

\bibitem{Cornella:2019uxs} 
  C.~Cornella, P.~Paradisi and O.~Sumensari,
  JHEP {\bf 2001}, 158 (2020)
  [arXiv:1911.06279 [hep-ph]].



\bibitem{Nie:1998dg} 
  S.~Nie and M.~Sher,
  Phys.\ Rev.\ D {\bf 58}, 097701 (1998)
  [hep-ph/9805376].

\bibitem{DiazRodolfo:2000yy} 
  R.~A.~Diaz, R.~Martinez and J.~A.~Rodriguez,
  Phys.\ Rev.\ D {\bf 64}, 033004 (2001)
  [hep-ph/0010339].
 
\bibitem{Iltan:2001nk} 
  E.~O.~Iltan and H.~Sundu,
  Acta Phys.\ Slov.\  {\bf 53}, 17 (2003)
  [hep-ph/0103105].
  
\bibitem{Wu:2001vq} 
  Y.~L.~Wu and Y.~F.~Zhou,
  Phys.\ Rev.\ D {\bf 64}, 115018 (2001)
  [hep-ph/0104056].

\bibitem{Assamagan:2002kf} 
  K.~A.~Assamagan, A.~Deandrea and P.~A.~Delsart,
  Phys.\ Rev.\ D {\bf 67}, 035001 (2003)
  [hep-ph/0207302].
  
\bibitem{Davidson:2010xv} 
  S.~Davidson and G.~J.~Grenier,
  Phys.\ Rev.\ D {\bf 81}, 095016 (2010)
  [arXiv:1001.0434 [hep-ph]].
  
\bibitem{Omura:2015nja} 
  Y.~Omura, E.~Senaha and K.~Tobe,
  JHEP {\bf 1505}, 028 (2015)
  [arXiv:1502.07824 [hep-ph]].
  
\bibitem{Omura:2015xcg} 
  Y.~Omura, E.~Senaha and K.~Tobe,
  Phys.\ Rev.\ D {\bf 94}, no. 5, 055019 (2016)
  [arXiv:1511.08880 [hep-ph]].

\bibitem{Iguro:2018qzf} 
  S.~Iguro and Y.~Omura,
  JHEP {\bf 1805}, 173 (2018)
  [arXiv:1802.01732 [hep-ph]].

\bibitem{Abe:2019bkf} 
  Y.~Abe, T.~Toma and K.~Tsumura,
  JHEP {\bf 1906}, 142 (2019)
  [arXiv:1904.10908 [hep-ph]].

\bibitem{Iguro:2019sly} 
  S.~Iguro, Y.~Omura and M.~Takeuchi,
  JHEP {\bf 1911}, 130 (2019)
  [arXiv:1907.09845 [hep-ph]].
  
\bibitem{Wang:2019ngf} 
  L.~Wang and Y.~Zhang,
  Phys.\ Rev.\ D {\bf 100}, no. 9, 095005 (2019)
  [arXiv:1908.03755 [hep-ph]].
  
\bibitem{Baek:2001kca} 
  S.~Baek, N.~G.~Deshpande, X.~G.~He and P.~Ko,
  Phys.\ Rev.\ D {\bf 64}, 055006 (2001)
  [hep-ph/0104141].

\bibitem{Heeck:2016xkh} 
  J.~Heeck,
  Phys.\ Lett.\ B {\bf 758}, 101 (2016)
  [arXiv:1602.03810 [hep-ph]].
  
\bibitem{Altmannshofer:2016brv} 
  W.~Altmannshofer, C.~Y.~Chen, P.~S.~Bhupal Dev and A.~Soni,
  Phys.\ Lett.\ B {\bf 762}, 389 (2016)
  [arXiv:1607.06832 [hep-ph]].

\bibitem{Iguro2020no1}
S.~Iguro, Kirtimaan A. Mohan and C.-P. Yuan,
arXiv:2001.09079 [hep-ph].

\bibitem{CMS:2017wua} 
  CMS Collaboration [CMS Collaboration],
  CMS-PAS-EXO-17-006.

\bibitem{Sirunyan:2018nwe} 
  A.~M.~Sirunyan {\it et al.} [CMS Collaboration],
  Phys.\ Lett.\ B {\bf 790}, 140 (2019)
  [arXiv:1806.05264 [hep-ex]].

\bibitem{Dev:2017ftk} 
  P.~S.~B.~Dev, R.~N.~Mohapatra and Y.~Zhang,
  Phys.\ Rev.\ Lett.\  {\bf 120}, no. 22, 221804 (2018)
  [arXiv:1711.08430 [hep-ph]].
  
\bibitem{Evans:2019xer} 
  J.~A.~Evans, P.~Tanedo and M.~Zakeri,
  JHEP {\bf 2001}, 028 (2020)
  [arXiv:1910.07533 [hep-ph]].

\bibitem{Abe:2010gxa} 
  T.~Abe {\it et al.} [Belle-II Collaboration],
  arXiv:1011.0352 [physics.ins-det].
\bibitem{Kou:2018nap} 
  E.~Kou {\it et al.} [Belle-II Collaboration],
  PTEP {\bf 2019}, no. 12, 123C01 (2019)
  [arXiv:1808.10567 [hep-ex]].
  
   \bibitem{Iguro2020no2}
  M.~Endo, S.~Iguro and T.~Kitahara,
  arXiv:2002.05948 [hep-ph].
  
  
  
\bibitem{Jho:2019cxq} 
  Y.~Jho, Y.~Kwon, S.~C.~Park and P.~Y.~Tseng,
  JHEP {\bf 1910}, 168 (2019)
  [arXiv:1904.13053 [hep-ph]].

 
\bibitem{Tanabashi:2018oca} 
  M.~Tanabashi {\it et al.} [Particle Data Group],
  Phys.\ Rev.\ D {\bf 98}, no. 3, 030001 (2018).
\bibitem{Ikado:2006un} 
  K.~Ikado {\it et al.} [Belle Collaboration],
  Phys.\ Rev.\ Lett.\  {\bf 97}, 251802 (2006)
  [hep-ex/0604018].
  
\bibitem{Hanagaki:2001fz} 
  K.~Hanagaki, H.~Kakuno, H.~Ikeda, T.~Iijima and T.~Tsukamoto,
  Nucl.\ Instrum.\ Meth.\ A {\bf 485}, 490 (2002)
  [hep-ex/0108044].

\bibitem{Alwall:2014hca} 
  J.~Alwall, R.~Frederix, S.~Frixione, V.~Hirschi, F.~Maltoni, O.~Mattelaer, H.-S.~Shao and T.~Stelzer {\it et al.},
  JHEP {\bf 1407}, 079 (2014)
  [arXiv:1405.0301 [hep-ph]].
  
\bibitem{Sjostrand:2006za} 
  T.~Sjostrand, S.~Mrenna and P.~Z.~Skands,
  JHEP {\bf 0605}, 026 (2006)
  [hep-ph/0603175].
  
   \bibitem{Alloul:2013bka} 
  A.~Alloul, N.~D.~Christensen, C.~Degrande, C.~Duhr and B.~Fuks,
  Comput.\ Phys.\ Commun.\  {\bf 185}, 2250 (2014)
  [arXiv:1310.1921 [hep-ph]].
 
\bibitem{Adachi:2019otg} 
  I.~Adachi {\it et al.} [Belle-II Collaboration],
  arXiv:1912.11276 [hep-ex].
 
 
 }}

\end{thebibliography}
\end{document}